\documentclass{appolb}
\def\BR{B}

\usepackage{amsmath}
\usepackage{amssymb}
\usepackage{graphicx}
\begin{document}

\title{False-vacuum decay and flaws in Frampton's model of the origin of life%
\thanks{Presented at the XLVI International Conference of Theoretical
  Physics {\em Matter To The Deepest: Recent Developments In Physics
    Of Fundamental Interactions}, Katowice, Poland, September 15-19, 2025.}%
}
\author{Andrzej Czarnecki and Jishnu Khanna
\address{Department of Physics, University of Alberta, Edmonton, Alberta, Canada}
}
\maketitle
\begin{abstract}
  We briefly review false-vacuum decay and examine a recent proposal
  by Frampton to model the origin of the first single-celled organism
  (SCO) as a phase transition between no-life and life vacua.  In his
  calculation the exponent $n$ entering the probability
  $P_{\rm SCO}\sim 10^{-n}$ has dimensions of inverse time: it is an
  energy barrier divided by the Planck constant, rather than a
  dimensionless tunnelling action. The resulting probability is
  mathematically ill-defined and does not determine a tunnelling
  rate. Apart from this dimensional issue, the assumed initial
  configuration, a toroidal structure made of long molecules, and its
  treatment in empty space are inconsistent with soft-matter physics
  and with the hot, collisional environment expected for prebiotic
  chemistry. Consequently, the claimed exponential suppression of
  biogenesis, and the inference that extraterrestrial life is likely
  absent, are not supported.
\end{abstract}

\section{Introduction}
\label{sec:intro}

Metastable, or false, vacua and their decay by bubble nucleation have
been studied in quantum field theory (QFT) since the 1970s. Lee and
Wick \cite{Lee:1974ma} and, more extensively, Voloshin, Kobzarev and
Okun (VKO) \cite{Kobzarev:1974cp} analysed the semiclassical
tunnelling of bubbles in a scalar theory with two minima of the
potential (see \cite{Okun:1994re} for the historical context). Their
work was extended and improved by Coleman
\cite{Coleman:1977py,Coleman:1985rnk}.  Related Euclidean
semiclassical methods were developed independently around the same
time by Stone \cite{Stone:1975bd,Stone:1976qh}.  Coleman and Callan
determined first quantum corrections \cite{Callan:1977pt}, while Linde
\cite{Linde:1977mm,Linde:1979ny,Linde:1980tt} developed the
corresponding analysis for the Weinberg-Salam model and for finite
temperature in cosmology (for a pedagogical presentation see
\cite{Linde:1990flp} and for recent work and references, see, e.g.,
\cite{Christie:2025knc}).  A common feature of all these treatments is
that the tunnelling probability is exponentially suppressed by a
Euclidean action, not by an energy barrier (see below).

In a recent paper \cite{Frampton:2025bzo} Frampton has proposed to use
the false-vacuum formalism to address the origin of life on Earth.  He
regards a prebiotic state without life as a false vacuum and a state
with life as the true vacuum, and models the appearance of the first
single-celled organism (SCO) as the nucleation of a bubble of the true
phase in a first-order transition.  Treating the problem as a
zero-temperature tunnelling process in vacuum, he arrives at a
probability
\begin{equation}
  P_{\rm SCO} \sim 10^{-n},
  \qquad
  n \simeq 9.3\times 10^{31},
  \label{eq:intro-PSCO}
\end{equation}
and, from the extreme smallness of $ P_{\rm SCO} $, concludes that
the origin of life is so improbable that it has almost certainly never
occurred elsewhere in the observable universe.  

The purpose of the present note is to examine Frampton's model as an
application of first-order phase-transition physics and to ask whether
his conclusion~\eqref{eq:intro-PSCO} is reliable.  We will argue that
it is not.  The problems fall into several classes: First, the
tunnelling exponent is being misused. In VKO, Coleman and Linde the
decay rate has the form $\Gamma/V \sim A e^{-\BR}$, where $\BR\sim \sigma^4/\varepsilon^3$ is a
dimensionless Euclidean action built from 
surface tension and energy-density difference and $A$ carries the
dimensions.  Frampton instead exponentiates a barrier energy
$E_{\mathrm{m}}\sim\sigma^3/\varepsilon^2$ and inserts the Planck
constant $h$ by hand, leading to an exponent $n$ which  has units of inverse time.  The central quantity
$P_{\rm SCO}\sim 10^{-n}$ is mathematically ill-defined: it
is the exponential of a dimensionful quantity.

Second, the model assumes that, before life, the Earth already
contains a planar rectangular sheet of long organic molecules that
behaves as a homogeneous cell membrane, rolls into a cylinder, and
closes into a torus.  This object is then treated as a false vacuum
which must tunnel to a spherical cell.  Such a planar sheet is a very
special, low-entropy configuration, and toroidal bubbles relax to
spheres by elasticity, not by quantum tunnelling.

Third, the calculation is carried out as if the SCO nucleated in
cold vacuum, whereas any plausible prebiotic environment is a
hot, dense aqueous medium with collisions and thermal noise.
Linde \cite{Linde:1977mm,Linde:1980tt} emphasised that in similar
environments the exponential suppression is tempered and
zero-temperature vacuum tunnelling is no longer the right description.

Fourth, it is not at all obvious that life automatically starts with
the change of topology from toroidal to spherical. There are many
lumps of organic matter with the topology of a sphere which are
perfectly dead. 

This note is organized as follows.
Section~\ref{sec:false} summarises the treatment of
thin-wall bubbles and the  Wentzel-Kramers-Brillouin (WKB) exponent
governing the vacuum decay.  
Section~\ref{sec:framptonmodel} describes
Frampton's model, with particular attention to the definition and
dimension of his exponent $n$. 
Section~\ref{sec:geometry} critiques  the assumed toroidal geometry
and its dynamics.
The conclusion is that the false-vacuum analogy in
Ref.~\cite{Frampton:2025bzo} does not support any claim about the
probability of life's emergence, let alone a statement about the
(non-)existence of extraterrestrial life.

\section{Metastable vacuum decay}
\label{sec:false}

False-vacuum decay can be described  as quantum tunnelling of
a scalar field from a metastable minimum of the
potential (false vacuum) to a lower minimum (true vacuum).  The decay
proceeds by nucleation of bubbles of true vacuum which then expand, 
converting the surrounding false vacuum.

\subsection{Thin-wall bubbles in a false vacuum}
\label{subsec:thinwall}

Consider a real scalar field $\varphi$ with a potential $U(\varphi)$
that has two local minima with different energy densities.  A
prototype is a double-well potential tilted so that the minimum at
$\varphi = \varphi_+$ is higher than  at $\varphi = \varphi_-$.

VKO consider a spherical bubble of true vacuum $\varphi_-$ of radius $R$ inside the
false vacuum.  In the thin-wall approximation, the range of $r$ where
$\varphi(r)$ varies significantly (the wall) is narrow in comparison with the
radius of the bubble $R$ (\cite{Linde:1990flp}, Section 5.2).  The bubble wall has a surface
tension~$\sigma$ (energy per unit area, resulting from $\varphi$ being
away from either minimum and from its gradient), and the difference of energy
densities between the two vacua is 
\begin{equation}
  \varepsilon \equiv \rho_{\text{false}} - \rho_{\text{true}} > 0.
\end{equation}
The energy of a static bubble of the lower-energy vacuum, of radius $R$, is 
\begin{equation}
  E(R) = 4\pi R^2\sigma - \frac{4\pi}3 R^3\varepsilon .
  \label{eq:thinwallE}
\end{equation}
The first term is the positive surface energy of the wall; the
second, the negative volume energy gain from replacing false by true
vacuum. 

Extremising \eqref{eq:thinwallE} with respect to $R$ gives the
radius at which $E(R)$ is maximum,
\begin{equation}
  R_m = \frac{2\sigma}{\varepsilon},
  \label{eq:Rm}
\end{equation}
and the height of the energy barrier,
\begin{equation}
  E_{\max} = E(R_m) = {16\pi \over 3} \frac{\sigma^3}{\varepsilon^2} .
  \label{eq:Emax}
\end{equation}
A bubble smaller than $R_m$ shrinks away; a bubble larger than
$R_m$ grows and converts the surrounding false vacuum into the true
one.

\subsection{WKB exponent and the vacuum decay rate}
VKO consider a zero-energy bubble of true vacuum, created by tunnelling
in the false vacuum. They find the probability of penetrating the
energy barrier to be proportional to $\exp(-\BR)$ with
\begin{equation}
  \label{eq:1}
  \BR = \frac{2}{\hbar}\int_0^{R_c} |p_R|\mathrm dR,
\end{equation}
where $p_R$ is the radial momentum of the wall and $R_c=3 \sigma/\varepsilon$ is the
radius at which $E(R)$ returns to zero \cite{Kobzarev:1974cp}. VKO
find, up to a dimensionless factor,
\begin{equation}
  \label{eq:2}
  \BR \sim {\sigma^4 \over \hbar c \varepsilon^3}.
\end{equation}
Here $\hbar=h/(2\pi)$ is the reduced Planck constant and $c$ is the speed of light.
In SI units $[\sigma]=\mathrm{J\,m^{-2}}$,
$[\varepsilon]=\mathrm{J\,m^{-3}}$ (see Eq.~\eqref{eq:thinwallE}), so
$[\sigma^4/\varepsilon^3]=\mathrm{J\,m}$, while
$[\hbar c]=(\mathrm{J\,s})(\mathrm{m\,s^{-1}})=\mathrm{J\,m}$.  Hence
$\BR \sim \sigma^4/(\hbar c\,\varepsilon^3)$ is dimensionless, as an exponent must be.

\section{Frampton's model of the origin of life}
\label{sec:framptonmodel}
In his 2025 paper \cite{Frampton:2025bzo}, Frampton models the origin
of life on Earth as a phase transition between two vacua: without
(false) and
with life  (true). 
\subsection{Summary of Frampton's argument}

Frampton assumes that long organic molecules  such as nucleic
acids assemble into a rectangular sheet which rolls to form a
cylinder, which subsequently closes on itself making a torus. 
Frampton identifies the torus-to-sphere change with the bubble
nucleation event.
The toroidal configuration is treated as a false vacuum, while a
spherical configuration is treated as
the stable true vacuum. Adapting the thin-wall bubble analysis of the
false-vacuum decay, he writes the energy of a spherical bubble of
radius $R$ as a sum of volume and surface contributions, see
Eq.~\eqref{eq:thinwallE}, with $\varepsilon$ interpreted as the
difference in volume-energy density between the two configurations,
\begin{equation}
  \label{eq:4}
  \varepsilon = \varepsilon_{\text{torus}}- \varepsilon_{\text{sphere}},
\end{equation}
which he estimates to be on the order of $10^{-3}$ in SI units, that
is joule per cubic metre. $\sigma$ is the surface tension of the wall,
estimated as $2\cdot 10^{-3}$ J/m$^2$. He finds $R_m$ as in Eq.~\eqref{eq:Rm}
$E_m =E(R_m) =16\pi\sigma^3/(3\varepsilon^2)$ as in  \eqref{eq:Emax}. He
assumes the probability of the  tunnelling to be
\begin{equation}
  \label{eq:5}
  P_{\text{tunnelling}} = \exp(-E_m) = 10^{-n},
\end{equation}
which defines the exponent $n$, with $n=E_m/\ln 10$. Inserting  a
factor of the Planck constant $h$ (Frampton uses $h$ rather than $\hbar$) in the denominator to
``restore the units from the use of $h=c=1$ natural units'', he finds 
\begin{equation}
  \label{eq:6}
  n ={ 16\pi\sigma^3 \over 3\ln 10\, h \varepsilon^2}.
\end{equation}
and claims $n=9.3\cdot 10^{31}$. This leads to the estimate of producing a
single-celled organism,
\begin{equation}
  \label{eq:7}
  P_{\rm SCO} \sim 10^{-n} \sim 10^{-9.3\cdot 10^{31}},
\end{equation}
a very small number. From the huge magnitude of the exponent he
concludes that the origin of life is so improbable that it is unlikely
ever to occur elsewhere in the visible universe.

\subsection{$n$ is not dimensionless}
The numerical estimate \eqref{eq:7} rests on treating $n$ as a pure
number. However, $n$ as defined in Eqs.~\eqref{eq:5}-\eqref{eq:6} is
not dimensionless. Obtained by dividing an energy by the Planck
constant, it has units of inverse time. Substituting numerical values quoted
in  \cite{Frampton:2025bzo}, one reproduces the magnitude $9.3\cdot
10^{31}$, but with units of inverse seconds. 

The source of this confusion is the use of the energy barrier
$E_m$ instead of the WKB exponent for the tunnelling bubble $\BR$
(see Eqs.~\eqref{eq:1}-\eqref{eq:2}). 
While $\BR$ is
dimensionless, the ratio  $E_m/h$ has units of inverse time.

Exponentials such as $10^{-n}$ are mathematically meaningful only if
the exponent is dimensionless. With $[n]=s^{-1}$ the main result $
P_{\rm SCO} $ in Eq.~\eqref{eq:7} is not well
defined. Consequently, the spectacularly small number in
Eq.~\eqref{eq:7}  has no physical content. 

\section{Implausibility of the torus-in-the-vacuum model}
\label{sec:geometry}

Aside from the issue with the tunnelling exponent, the assumed
degrees of freedom and environment in Ref.~\cite{Frampton:2025bzo} are
implausible for prebiotic chemistry.

\subsection*{Membrane composition}

Frampton assumes that long organic molecules such as nucleic acids
can assemble into a rectangular sheet, which then rolls into
a hollow cylinder and closes into a torus.  This starting point is already
highly non-generic: long, flexible polymers are not, by themselves,
likely membrane-forming materials.  In contrast, much of mainstream
origin-of-life work focuses on membranes made of simpler
molecules (for example fatty acids), which have a water-attracting head
and a water-repelling tail and can spontaneously self-assemble in water
into closed bubbles called vesicles \cite{Szostak:2001}.

Laboratory work supports the plausibility of such compartments:
fatty-acid-based vesicles can be stable over a wide temperature range
(up to near boiling) and can retain encapsulated short nucleic-acid strands
\cite{MansySzostak2008}.  Recent prebiotic-chemistry experiments also
demonstrate routes from relatively simple building blocks to
membrane-forming lipids that self-assemble into protocell-like vesicles
\cite{Cho2025NatChem}.  A concise review of protocell-based approaches and
the idea that membranes and primitive genetic polymers likely co-evolved is
given in Ref.~\cite{Schrum2010}. For an accessible discussion in the context of extraterrestrial
life see the recent book~\cite{livio2024earth}.

\subsection*{Toroidal shapes relax classically}

Now suppose that a toroidal membrane is somehow formed.  Let it be characterized
by two radii $a$ and $b$ with $a>b$.  Such a structure relaxes toward a sphere
through classical dynamics; no quantum tunnelling is required.
Intuitively, decreasing the large radius $a$ at approximately fixed volume reduces
the surface area, and the system can lower its free energy by shrinking the torus.
This is consistent with experimental and theoretical studies of toroidal droplets
and their shrinking instability \cite{toricInstab,PhysRevLett.102.234501,yao2011shrinking}.

\subsection*{Thermal environment and the relevant energy scale}

Frampton treats the prebiotic toroidal structure as if it
existed in vacuum.  However, life has likely appeared in messy
environments, so membrane shapes are influenced by thermal
fluctuations and collisions.  Frampton argues that the temperature $T$
even near an oceanic thermal vent is low (when converted into energy
using the Boltzmann constant $k_B \simeq 10^{-4}$ eV/K) in comparison
with the hydrogen ionization energy $E_H = 13.6$ eV
\cite{Frampton:2025bzo}.  This argument is flawed since one can
significantly distort a membrane with much less energy than the energy
$\sim E_H$ that is needed to turn it into plasma.

The ionization scale $E_H$
is an electronic excitation energy, whereas the energy scale controlling membrane
deformations is set by soft-matter elasticity.  Membrane deformations are described by the Helfrich curvature elasticity
\cite{Helfrich1973}, with a typical bending rigidity for lipid bilayers of
order $\kappa \sim 20\,k_BT$ \cite{kozlov2015membrane}.  At $T\simeq 373\,\mathrm{K}$
one has $k_B T \simeq 0.032\,\mathrm{eV}$, hence $\kappa \sim 0.6\,\mathrm{eV}$,
many times smaller than $E_H$.  Therefore comparing $T$ to $13.6\,\mathrm{eV}$ does
not constrain membrane-scale distortions in the way implied in Ref.~\cite{Frampton:2025bzo}.

\section{Conclusions}
The goal of this note has been to compare Ref.~\cite{Frampton:2025bzo}
with the standard semiclassical description of false-vacuum decay.
Section \ref{sec:false} recalled that in that framework, the decay of
metastable vacuum is governed by a dimensionless exponent $\BR$, with
the nucleation rate proportional to $\exp(-\BR)$. Barrier energies
such as $E_m \sim \sigma^3 /\varepsilon^2$ appear only as ingredients
of $\BR$, never alone in the exponent.

Ref.~\cite{Frampton:2025bzo} instead assigns the tunnelling probability the form
$P_{\rm SCO}\sim 10^{-n}$ with $n$ proportional to a barrier energy divided by
Planck's constant.  The resulting exponent has dimensions of inverse time, so
$10^{-n}$ is not a mathematically meaningful probability.  Consequently the
numerical estimate $P_{\rm SCO}\sim 10^{-9.3\times 10^{31}}$ has no physical
content and cannot be used to infer the rarity of biogenesis or to constrain the
existence of extraterrestrial life.

Even setting aside this dimensional problem, the setup assumed in
Ref.~\cite{Frampton:2025bzo} is mismatched to prebiotic chemistry.  The proposed
initial state, a toroidal membrane built from long polymers, is not a likely
membrane-forming configuration; toroidal shapes relax classically; and the
relevant energy scales for membrane deformations are
set by soft-matter elasticity (e.g.\ $\kappa\sim 20\,k_BT\simeq 0.6$
eV), not by electronic
ionization energies such as $13.6$ eV.  For these reasons, applying
false-vacuum tunnelling formulas in vacuum does not provide a quantitative basis
for the conclusions drawn in Ref.~\cite{Frampton:2025bzo}.

The overall message is optimistic: although the origin of life
remains a mystery, its probability is not obviously prohibitively
small. Perhaps we are not alone in the universe.

\subsection*{Acknowledgments}
This research was funded
by Natural Sciences and Engineering Research Canada (NSERC). We are
grateful to
Jan Czarnecki and Michael T.~Woodside for helpful discussions.  We
thank Fadilah Abolade for collaboration at an early stage of this project.
%\bibliographystyle{/Users/czar/Dropbox/pro/Tables/Archive/BibTeX/andrzej}
%\bibliography{/Users/czar/Dropbox/pro/Tables/Archive/ac}

\end{document}